\documentclass[preprint,10pt]{elsarticle}
\usepackage{amsmath}
\usepackage{enumerate}
\usepackage{array}
\usepackage{graphicx}
\usepackage[footnotesize,bf]{caption}
\usepackage{subcaption}
\usepackage{amsfonts}

\journal{Physica A}
\begin{document}
\begin{frontmatter}
\title{Complex scale-free networks with tunable power-law exponent and clustering}
\author{E.R. Colman}
\ead{Ewan.Colman@brunel.ac.uk}
\author{G.J. Rodgers}
\address{Department of Mathematical Sciences, Brunel University, Uxbridge, Middlesex UB8 3PH, U.K.}
\begin{abstract}
We introduce a network evolution process motivated by the network of citations in the scientific literature. In each iteration of the process a node is born and directed links are created from the new node to a set of target nodes already in the network. This set includes $m$ ``ambassador'' nodes and $l$ of each ambassador's descendants where $m$ and $l$ are random variables selected from any choice of distributions $p_{l}$ and $q_{m}$. The process mimics the tendency of authors to cite varying numbers of papers included in the bibliographies of the other papers they cite. We show that the degree distributions of the networks generated after a large number of iterations are scale-free and derive an expression for the power-law exponent. In a particular case of the model where the number of ambassadors is always the constant $m$ and the number of selected descendants from each ambassador is the constant $l$, the power-law exponent is $(2l+1)/l$. For this example we derive expressions for the degree distribution and clustering coefficient in terms of $l$ and $m$. We conclude that the proposed model can be tuned to have the same power law exponent and clustering coefficient of a broad range of the scale-free distributions that have been studied empirically.
\end{abstract}

\end{frontmatter}
\section{Introduction}
Networks, in recent years, have become ubiquitous in the modelling of complex systems. Many fields of study, including for example biology \cite{bio}, economics \cite{eco}, and epidemiology \cite{epidemic}, employ network based models to mimic the large numbers of agents that interact in the systems they study. It is often found that with the aid of an appropriate network model, the the macroscopic behaviour of a complex system can be reproduced with very few assumptions being made about the constituent agents themselves. Typically the system being modelled will be reduced to a set of vertices and a set of vertex pairs called edges. Vertices and edges may represent things like web pages and the hyper-links between them \cite{www}, people and their friendships \cite{friends}, or transport hubs and the transport links between them \cite{transport}. In many cases the structure of the network exhibits non-trivial statistical properties such as a high level of clustering, short average path lengths and small numbers of highly connected vertices. An example that is frequently used to illustrate these properties is the network of citations in the scientific literature. Here we present a stochastic model that closely resembles this network but also has the potential to have many other applications.
\newline

A crude but arguably effective measure of the worth of a scientific paper is the number citations made to it from other existing scientific articles. Empirical studies have shown that the number of articles with $k$ citations (i.e. cited by $k$ other articles) is proportional to $k^{-3}$ \cite{price,redner}. This distribution has certain properties we might expect, namely that the vast majority of papers written have few citations, creating little or no impact on future research, whereas a very small number of papers are extremely significant and have a very large number of citations.
By modelling each paper as a node (vertex) and drawing directed edges from each paper to the papers it cites, it has been shown that the correct degree distribution is reproduced using \emph{preferential attachment}; the process of creating nodes sequentially and linking them to nodes selected randomly with probability proportional to their degree (originally discussed in \cite{price76} although the term was coined later in \cite{Albert}). The implication of this result is that authors of scientific articles are more likely to choose to cite articles that are already well cited rather than ones that have few or no citations. The attractiveness of highly connected nodes can be explained by a number of processes for example \emph{redirection} \cite{organisation}, where nodes are selected randomly and a link is formed between one of its neighbours and a new node, and random walk models \cite{vazquez,saramaki,evans} where the new node is linked to the nodes occupied by random walkers on the network.
\newline

\begin{figure}[t]
 \centering
        \begin{subfigure}[b]{0.4\textwidth}
                \centering
                \includegraphics[width=\textwidth]{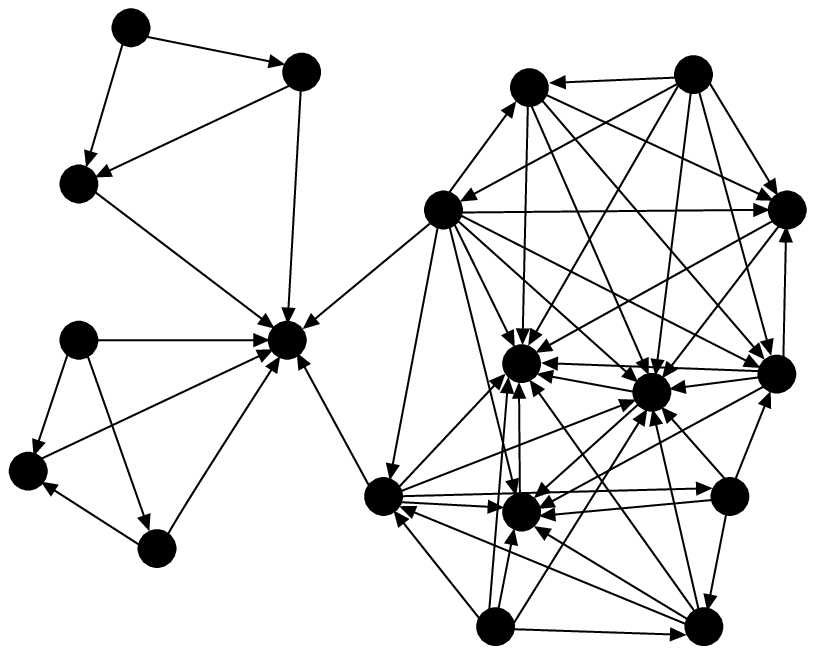}
                \caption{The reference graph of a real paper.}
                \label{realSub}
        \end{subfigure}
        \qquad
        \begin{subfigure}[b]{0.4\textwidth}
                \centering
                \includegraphics[width=\textwidth]{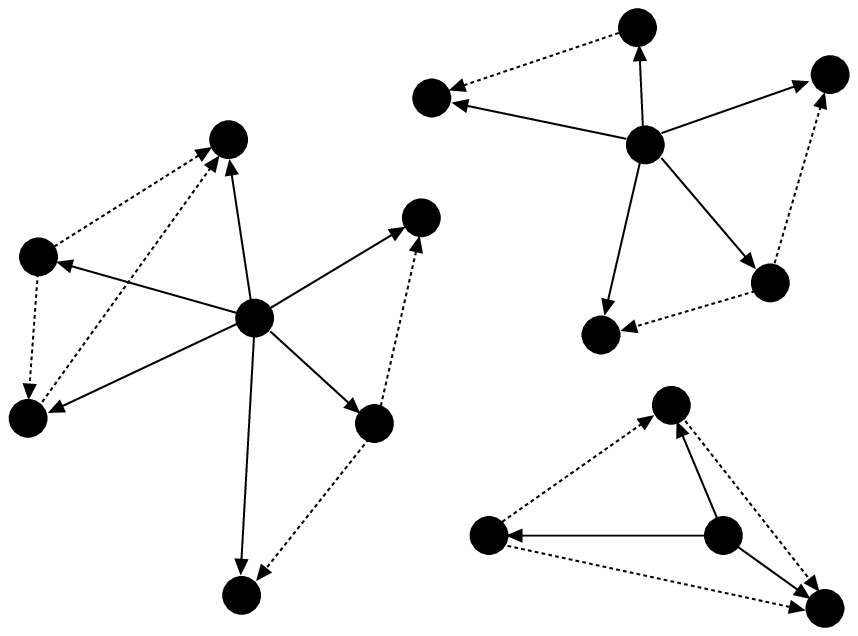}
                \caption{A typical set of nodes selected in one iteration of the model.}
                \label{modelSub}
        \end{subfigure}
\caption{\ref{realSub} shows the sub-network know as the reference graph of a real paper (taken from \cite{empirical}) the nodes represent the papers cited by that paper and the edges represent the citations between them, it is highly clustered and many of the nodes are descendants of others. \ref{modelSub} shows the a typical sub-network structure of nodes that the proposed model links to, in this case three nodes are selected initially and 3, 5, and 6 of their descendants are also selected, the dotted lines represent the other links between descendants.}
\label{fig:figure0}
\end{figure}

 There is a growing literature offering more accurate representations of the way in which the citation network develops, much of this work can be found in the fields of Scientometrics, Bibliometrics, Informetrics and Webometrics \cite{bornmann,SciReview,borner}. A significant amount has been written concerning models that not only agree with the empirical data regarding degree distributions but also agree with other properties, for example in \cite{borner} the evolution of the citation network model is motivated by a coupling with the network of co-authors, other models account for the effect of time on the probability of receiving a citation \cite{Wang, time}. The model in \cite{holme} introduces tunable clustering (quantified by the clustering coefficient \cite{statmech}) by extending the preferential attachment model with an additional Triad Formation (TF) step. For each node that is introduced to the network, a node is selected by preferential attachment and linked to, then each neighbouring node is selected with probability $p$ and also linked to from the new node, resulting in a triangle (triad) of edges. The forest fire model described in \cite{kleinberg} extends the Triad Formation model by selecting multiple neighbours of the initially selected node, the process continues by then linking to a number of the neighbours of those neighbours and so on, at each stage a random variable from the binomial distribution determines the number of neighbours selected. In \cite{empirical} the forest fire model, along with other models that attempt to mimic the network of citations in even greater detail, is tested against empirical data. The authors also examine the way the articles cited by any one paper, call it $i$, relate to one another forming a sub network called a reference graph of $i$ (see Fig.\ref{realSub}). They observed that a clique structure is prevalent, i.e small groups of nodes that all link to each other, and incorporated this finding into their own model.
 \newline

 Much of the literature suggests that the high levels of clustering found in citation networks is a consequence of each author's choice to cite papers that are found in the bibliographies the other papers they cite. This has been observed empirically \cite{simkin2002read}, and modelled using a TF process where the initial nodes are selected randomly (rather than preferentially) \cite{Simkin}. A power-law degree distribution was found with an exponent that varies depending on the Triad Formation probability $p$, however, this model does not exhibit the exponential out-degree distribution observed in the data \cite{vazquez2001statistics}.
\newline

The models mentioned above and those considered in this paper belong to the class of evolving directed clustered scale-free networks that have applications beyond citation networks, the world-wide web being another well studied example. In these models the distributions of in-degree and out-degree are treated separately, often driven by a preferential linking mechanism where the probability of adding an edge from a node $i$ is proportional to the out-degree of $i$, similarly the probability that the link will end at node $j$ is proportional to the in-degree of $j$ \cite{Tadic}. Correlations between the in-degree and out-degree of nodes in such networks have been shown to emerge \cite{Rodgers}. A detail of citation networks that makes analysis substantially easier is that the out-degree of a node $i$ is fixed from the moment it is created. Consequently the evolution of the out-degree distribution can be disregarded, moreover we can control the out-degree distribution through an appropriate parametrization and ultimately answer the question of how the distribution of bibliography sizes affects the topology of the network.
\newline

In this paper we introduce a variant of the TF model that uses a different parameter set to those previously studied. Using the distributions for the number of initial citations and the number of copied citations (which together give the out-degree distribution) as parameters, we show that the networks created by this process may have power-law in-degree distribution with any exponent greater than $2$, and a clustering coefficient that ranges between $0$ and $1$. In Section \ref{description} we describe the stochastic process that iteratively grows the network. We describe a simple case of the model in Section \ref{mean} and solve for the power law exponent of the in-degree distribution in terms of two input parameters. In Section \ref{general} we formulate an expression for the in-degree distribution in terms of two input probability distributions then in Section \ref{clustering} we find an expression for the clustering coefficient for the model in Section \ref{mean}. In Section \ref{numerics} we present numerical results that confirm the results of Sections \ref{mean} and \ref{general}. In Section \ref{remarks} we discuss the strengths of our models and suggest how this work might be continued.
\section{The model}
\label{description}
Starting from a finite random network, at each iteration a node $j$ is introduced and directed links are formed between $j$ and a set of nodes that already exist in the network. Letting $p_{l}$ and $q_{m}$ be the probability distributions of the discrete random variables $l$ and $m$, links are formed by the following process:
\begin{enumerate}
\item The value $m$ is selected with probability $q_{m}$ and steps $2$ and $3$ are repeated a $m$ times.

\item The value $l$ is selected with probability $p_{l}$, a node $i$ in the network is randomly selected from those which have out-degree $l$ or greater, the edge $j\rightarrow i$ is added. Borrowing the terminology used in \cite{kleinberg} we will refer to $i$ as an ``ambassador''.

\item $l$ of $i$'s descendants are randomly selected and directed edges are added from $j$ to each of these.
\end{enumerate}
We are primarily interested in expressing the degree distributions for both incoming and outgoing edges and the clustering coefficient of the network as the number of iterations grows very large in terms of $p_{l}$ and $q_{m}$ ($l,m\in \mathbb{N}$). In the next section we solve for a simplified model where $l$ and $m$ are fixed (i.e. $p_{r}=\delta_{rl}$ and $q_{r}=\delta_{rm}$), we present the general solution in the section that follows.
\section{Attachment to $m$ random nodes and $l$ of each of their descendants}
\label{mean}
\begin{figure}[t]
  \centering
 \includegraphics[width=0.3\textwidth]{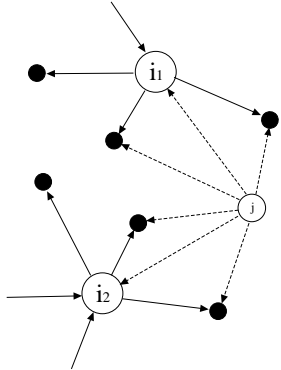}
  \caption{The new node $j$ attaches to 2 randomly selected nodes $i_{1}$ and $i_{2}$ as well as 2 randomly descendants shown here in black, the dashed lines represent the new edges that are added in this iteration whereas the solid ones were added previously. This example illustrates one iteration in the growth process when $m=2$ and $l=2$.}
  \label{fig:figure1}
\end{figure}
 We examine the network generated by the process described in Section \ref{description} when $p_{r}=\delta_{rl}$ and $q_{r}=\delta_{rm}$, in this section we derive the degree distribution of this network. In this simplified model the growth of the network depends only on the fixed values $l$ and $m$, thus the process can be described concisely as follows; in each iteration, $m$ ambassador nodes are randomly selected, $l$ descendants of each ambassador are also selected, then a new node $j$ is attached to each of the selected nodes (see Fig.\ref{fig:figure1}). We are interested in calculating the probability $P(k)$ of finding a node with in-degree $k$, in the citation model this represents the proportion of articles that are cited by $k$ other papers. Let $N$ be the total number of nodes, $N$ increases by $1$ with each iteration and every node has an out-degree of $m(l+1)$, the number of edges as $N$ grows large is $E=m(l+1)N$. Consider a typical node $i$ with in-degree $k$. There are two possible events which may cause the degree of $i$ to increase to $k+1$: $i$ can either be selected as one of the $m$ ambassador nodes, or it can be selected as a descendant of another node $j$. In any given iteration, $i$ will be selected as an ambassador with probability $m/N$. Alternatively $j$ will be selected as an ambassador with probability $m/N$ and then $i$ will be one of the $l$ selected descendants of $j$ with probability $l/m(l+1)$. Let $P(i;k)$ denote the probability that in one iteration a node $i$ with degree $k$ is selected. Since there are $k$ potential ancestors to $i$,
\begin{equation}
P(i;k)=\frac{m}{N}\left(1+\frac{lk}{m(l+1))}\right).
\end{equation}
It is possible for the same node to be selected two or more times in one iteration, for example if two of the selected ambassador nodes are a distance of one or two edges from each other. Since this possibility becomes less likely as $N$ increases we do not account for it in our calculations. Let $I_{k}$ be the number of nodes with in-degree $k$.  For $k\geq 1$, $I_{k}$ changes over time according to the rate equation
\begin{equation}
\label{rate1}
\frac{\partial I_{k}}{\partial N}=\frac{m}{N}\left[\left(1+\frac{l(k-1)}{m(l+1)}\right)I_{k-1}-\left(1+\frac{lk}{m(l+1)}\right)I_{k}\right].
\end{equation}
The first term on the right hand side accounts for the creation of a node of in-degree $k$ that occurs when one of the new edges attaches to a node of in-degree $k-1$, the second term accounts for the destruction of a node of in-degree $k$ when it is attached to by one of the new edges. For $k=0$ the rate equation is
\begin{equation}
\label{rate2}
\frac{\partial I_{0}}{\partial N}=1-\frac{m}{N}I_{0}.
\end{equation}
We are interested in finding $P_{\text{in}}(k)$ the probability of a node having in-degree $k$ when $N$ is very large. By assuming $P(k)$ grows linearly with $N$ when $N$ is large, we substitute $I_{k}=NP_{\text{in}}(k)$ into Eq.(\ref{rate1}) to find
\begin{eqnarray}
\left(\frac{l+1+m(l+1)}{l}+k\right)P_{\text{in}}(k)&=&\left(\frac{m(l+1)-l}{l}+k\right)P_{\text{in}}(k-1)
\end{eqnarray}
for $l\neq 0$. From Eq.(\ref{rate2}) we also find
\begin{equation}
P_{\text{in}}(0)=\frac{1}{1+m}
\end{equation}
and thus the in-degree distribution is expressed
\begin{equation}
\label{gammas}
P_{\text{in}}(k)=\frac{1}{m+1}\frac{\Gamma[(2l+1+m(l+1))/l]\Gamma[k+m(l+1)/l]}{\Gamma[m(l+1)/l]\Gamma[k+(2l+1+m(l+1))/l]}.
\end{equation}
For large enough values of $k$, $P_{\text{in}}(k)$ has power-law form
\begin{equation}
\label{exponent1}
P_{\text{in}}(k)\sim k^{-\gamma} \text{ where } \gamma=\frac{2l+1}{l}.
\end{equation}

\section{Solution to the general model}
\label{general}
\begin{figure}[t]
  \centering
 \includegraphics[width=0.3\textwidth]{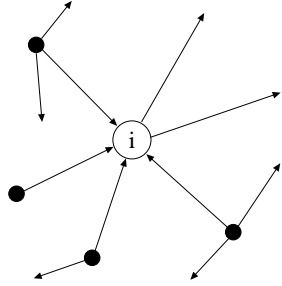}
  \caption{At each time-step $i$ can be selected as the initial node, alternatively one of the black nodes may be selected and then $i$ is selected as one of its descendants.}
  \label{fig:figure2}
\end{figure}
Let $P_{\text{out}}(s)$ denote the proportion of nodes in the network that have out-degree $s$. Note that at the time of its creation, the out-degree of a node is fixed and, unlike its in-degree, does not change over time. Therefore, for sufficiently large networks, $P_{\text{out}}(s)$ is equal to the probability of creating a node with out-degree $s$ within a single iteration. This can be written
\begin{equation}
\label{outdegree}
P_{\text{out}}(s)=\sum_{n=1}^{\infty}q_{n}P\left(\sum_{i=1}^{n}x_{i}=s\right)
\end{equation}
where the $x_{i}$ are integer random variables that equal $l+1$ with probability $p_{l}$.
\newline
To calculate the in-degree distribution we again construct a rate equation from the probability that the degree of a typical node $i$ will increase in one iteration. Let $T_{l}$ be the number of nodes have out-degree greater or equal to $l-1$, for the node $i$ to be randomly selected as the ambassador in step $2$ it must be one of these nodes. The probability that this is the case, multiplied by the probability that $i$ is the one node randomly selected from the $T_{l}$ nodes available, forms the probability that $i$ is the ambassador given that $l$ is the number of descendants chosen in step $2$. Summing over all values of $l$ returns $P_{a}(i)$ the probability that any node $i$ is selected as an ambassador, thus
\begin{eqnarray}
\label{emb}
P_{a}(i)&=&\sum_{l=1}^{\infty}p_{l}\frac{T_{l}}{N}\times\frac{1}{T_{l}}\nonumber\\
&=&\frac{1}{N}
\end{eqnarray}
Suppose $i$ has in-degree $k$, and that as per step $2$ only nodes with out-degree $l$ or greater can be selected as an ambassador. Suppose also that the ambassador is an ancestor of $i$ and has out-degree $s$ where $s\geq l$ (see Fig.\ref{fig:figure2}). The expectation of the number of nodes that satisfy these conditions is $kP_{\text{out}}(s)$. The probability that each one is selected is $1/N$ given by Eq.(\ref{emb}). Once selected, the probability that of the $s$ descendants $i$ is one of those selected in Step $3$, is $l/s$. Taking the product and summing over all $l$ and all possible values of $s$ returns $P_{d}(i;k)$ the probability that any node $i$ with degree $k$ is selected as a descendant, therefore
\begin{equation}
\label{des}
P_{d}(i;k)=\frac{k}{N}\Phi
\end{equation}
where
\begin{equation}
\Phi(p,q)=\sum_{l=1}^{\infty}\sum_{s=l}^{\infty}\frac{lp_{l}P_{\text{out}}(s)}{s}.
\end{equation}
The probability of a node $i$ with degree $k$ being linked to during step $2$ or $3$ of the process is $P_{a}(i)+P_{d}(i;k)$. Summing again over all possible values of $m$, the probability that the degree of $i$ will increase by $1$ during any iteration is
\begin{eqnarray}
P(i;k)&=&\frac{\langle m \rangle}{N}(1+k\Phi)
\end{eqnarray}
where
\begin{equation}
\langle m \rangle=\sum_{m=1}^{\infty}mq_{m}.
\end{equation}
The associated rate equation is constructed in exactly the same way as Eq.(\ref{rate1}), thus
\begin{equation}
\label{rate3}
\frac{\partial I_{k}}{\partial N}=\frac{\langle m \rangle}{N}\left[\left(1+(k-1)\Phi\right)I_{k-1}-\left(1+k\Phi\right)I_{k}\right].
\end{equation}
Letting $P_{\text{in}}(k)=I_{k}/N$ be the proportion of nodes that have in-degree at large $N$, Eq.(\ref{rate3}) becomes
\begin{equation}
\left(\frac{1+\langle m \rangle}{\langle m\rangle\Phi}+k\right)P_{\text{in}}(k)=\left(\frac{1-\Phi}{\Phi}+k\right)P_{\text{in}}(k-1).
\end{equation}
The rate equation for $I_{0}$ solves to find $P(0)=1/(1+\langle m \rangle)$ and thus
\begin{equation}
\label{gammas2}
P_{\text{in}}(k)=\frac{1}{\langle m\rangle+1}\frac{\Gamma[(1+\langle m\rangle)/\langle m\rangle\Phi]\Gamma[k+(1-\Phi)/\Phi]}{\Gamma[(1-\Phi)/\Phi]\Gamma[k+(1+\langle m\rangle)/\langle m\rangle\Phi]}.
\end{equation}
For large values of $k$, $P_{\text{in}}(k)$ has a power-law form
\begin{equation}
\label{exponent1}
P_{\text{in}}(k)\sim k^{-\gamma} \text{ where } \gamma=1+\frac{1}{\langle m\rangle\Phi}.
\end{equation}
\section{Clustering}
\label{clustering}
\begin{figure}[h]
  \centering
 \includegraphics[width=1\textwidth]{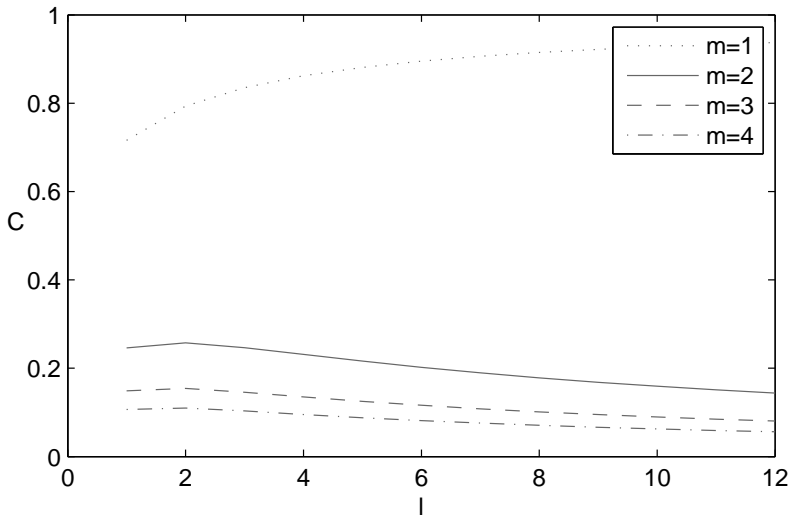}
  \caption{Using the formulae in Section \ref{clustering} the average clustering coefficient is plotted on the vertical axis for the first $4$ values of $m$, against $l$ on the horizontal axis. The clustering tends to zero as $l$ grows large for all values of $m$ with the exception of $m=1$.}
  \label{fig:figure6}
\end{figure}

The clustering coefficient of a node $i$ is defined as the the number of edges between the neighbours of $i$ divided by the number of pairs of nodes from the neighbours of $i$. If node $i$ has $d$ neighbours (ancestors and descendants) then this is
\begin{equation}
\label{cluster}
C_{i}=\frac{2E_{i}}{d(d-1)}
\end{equation}
where $E_{i}$ is the number of edges between the neighbours of $i$. Let $E(k)$ be the the expectation of $E_{i}$ when $i$ has in-degree $k$, also let $\Theta(k)$ be the expectation of the number of times $i$ has been selected as the ambassador node during step 2 of any previous iteration. From equations (\ref{emb}) and (\ref{des}) we see that the $k$th edge is $(k-1)\Phi$ times more likely to be added as a result of one of $i$'s neighbours being an ambassador rather than $i$ being selected as an ambassador itself, so
\begin{equation}
\label{theta}
\Theta(k)=\sum_{i=0}^{k-1}\frac{1}{1+i\Phi}.
\end{equation}
We find $\bar{C}$, the mean of $C_{i}$ over all nodes $i$ in the network we studied in Section \ref{mean} where $p_{r}=\delta_{rl}$ and $q_{r}=\delta_{rm}$. This is the sum over all $k$ of the product of $P_{\text{in}}(k)$ given by Eq.(\ref{gammas}), and $C(k)$ the expectation of the clustering of a node of degree $k$. The contribution to $E_{i}$ made by each neighbour $j$ of $i$ depends on the way in which the link was originally created, there are four cases to be considered. The first is where the link $i\rightarrow j$ was added when $i$ was introduced to the network, in this case the expected contribution to $E_{i}$ is the number of edges in the reference graph of $i$ (see Fig.\ref{fig:figure0}). The second case is where $i$ was selected as an ambassador and $l$ edges are added to $E_{i}$. In the third case $D_{a}$ edges are counted for those that were added when $i$ was selected as a descendant of the ambassador $j$ where the link from $j$ to $i$ was  originally formed when $i$ was selected as an ambassador in a previous iteration. Lastly, $D_{d}$ edges are counted for those that were added when $i$ was selected as a descendant of the ambassador $j$ where the link from $j$ to $i$ was formed when $i$ was originally selected as a descendant. Then
\begin{equation}
\label{expk}
E(k)=E(0)+\Theta l+(k-\Theta)\left[\frac{\Theta}{k} D_{a}+\frac{1-\Theta}{k}D_{d}\right].
\end{equation}
When a node $i$ is added to the network, $E_{i}$ includes the edges between each ambassador node and its descendants as well as the edges between those descendants. The probability that an edge exists between two descendants of the same node is $C(0)$ so the expectation of $E_{i}$ is
\begin{equation}
E(0)=m\left(l+\binom{l}{2}C(0)\right).
\end{equation}
Combining this with Eq.(\ref{cluster}) and solving gives
\begin{equation}
\label{C0}
C(0)=\frac{2l}{(m-1)(l+1)^{2}+2l}.
\end{equation}
In the instance where an ambassador node $j$ is selected and $i$ is linked to as one of $j$'s descendants, the new node will link to a further $l-1$ neighbours of $m(l+1)-1$ possible descendants of $j$, those that are also neighbours of $i$ will be counted in $E_{i}$. If $j$ originally formed a link with $i$ by selecting $i$ as an ambassador, then $l$ of $i$'s descendants are also descendants of $j$, hence the expectation of the number of neighbours of $i$ that are linked to is
\begin{equation}
\label{da}
D_{a}=\frac{l(l-1)}{m(l+1)-1}.
\end{equation}
If $j$ originally formed a link with $i$ by selecting $i$ as the descendant of some other node, the expected number of of links between $j$ and any of $i$'s neighbours is $1+(l-1)C(0)$ so the expectation of the number of $i$'s neighbours linked to is
\begin{equation}
\label{dd}
D_{d}=\frac{[1+(l-1)C(0)](l-1)}{m(l+1)-1}.
\end{equation}
Combining Equations (\ref{cluster}), (\ref{theta}), (\ref{expk}), (\ref{C0}), (\ref{da}) and (\ref{dd}) gives an expression for the clustering of a node of in-degree $k$ in terms of $m$ and $l$ ($l,m\geq 1$), multiplying by $P_{\text{in}}(k)$ given by Eq.(\ref{gammas}) and summing over all $k$ gives the mean clustering for the entire network. The clustering coefficient tends to $0$ as $m$ grows large. As $l$ grows large the clustering also tends to $0$ except when $m$ is equal to one, in which case it tends to $1$ (see Fig.\ref{fig:figure6}).

\section{Numerical results}
\label{numerics}
\begin{figure}[h]
  \centering
 \includegraphics[width=0.6\textwidth]{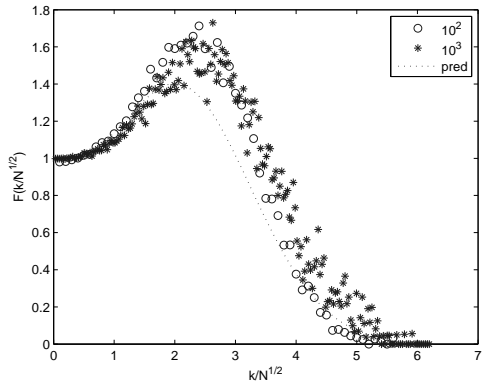}
  \caption{The correction function $F(\xi)$ in Eq.(\ref{correction}) for the simplest case of the model. Here $m=1$, $l=1$ and an initial condition of three nodes, one with out-degree $2$ connected to the other two. The dashed line represents the analytical prediction from \cite{finiteness}.}
  \label{fig:figure8}
\end{figure}

It should be emphasized that the results found in previous sections are mean field approximations as $N$ tends to infinity, it therefore cannot immediately be assumed that the derived results will be a fair description of any individual network grown following the proposed process. We consider the following:
\begin{enumerate}
\item Correlations between out-degree of a node and the in-degree of its descendants. Specifically in Eq.(\ref{des}) where it is assumed that the out-degree of the neighbours of node $i$ (i.e the black nodes in Fig.(\ref{fig:figure2})) follow the distribution $P_{\text{out}}(s)$ regardless of the in-degree of $i$. In reality this might not be the case; imagine, for example, a node $j$ with relatively large out-degree and $i$ as one of its descendants, selecting $j$ as an ambassador in future iterations is relatively unlikely to result in selecting $i$ again unless the new node also has large out-degree (more specifically a large value of $l$ in step $2$ of the iteration), so the expectation is for $i$ to have few ancestors each with large out-degree (the opposite is true if the out-degree of $j$ is small). The effect of this has not been considered analytically, instead we show numerically that in practice there is no significant deviation from the mean field result.

\item Finite size corrections. For finite networks of size $N$, the existence of a largest degree $k_{\text{max}}$ means the power-law degree distribution must fail around the largest values of $k$. These effects have been investigated for particular classes of preferential attachment based network \cite{evans,finiteness}. Once the asymptotic mean field solution $P(k)$ is known, the solution to average degree distribution on a network of size $N$ is
\begin{equation}
\label{correction}
N_{k}(N)\simeq NP(k)F(\xi) \text{ where } \xi=k/k_{\text{max}}.
\end{equation}
From the generated data (discussed below) we observe in Fig.\ref{fig:figure8} a similar form of scaling function $F(\xi)$ as observed in \cite{evans,finiteness}.
The function $F(\xi)$ can be derived by considering the average of all possible values of $N_{k}(N)$ for every $N$ starting from an initial value for $N_{1}(1)$, under the specific circumstances however, the initial conditions must be chosen carefully for each possible choice of our parameters. It is impractical to derive $F(\xi)$ for every possibility here, instead we show that the model passes a suitable goodness-of-fit test even when finite size effects are neglected.
\end{enumerate}

\begin{figure}[h]
 \centering
        \begin{subfigure}[b]{0.45\textwidth}
                \centering
                \includegraphics[width=\textwidth]{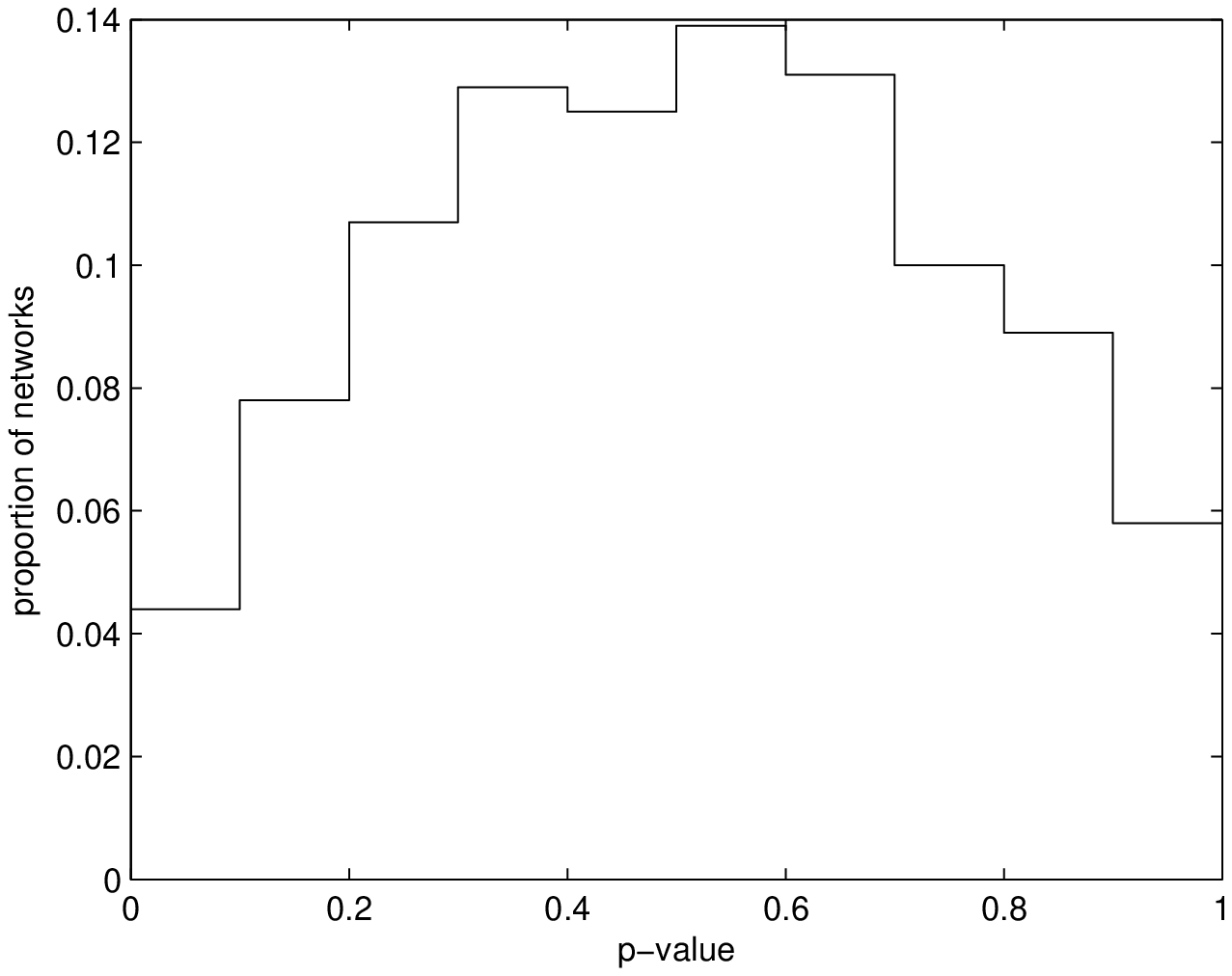}
                \label{figure8a}
        \end{subfigure}
        \qquad
        \begin{subfigure}[b]{0.45\textwidth}
                \centering
                \includegraphics[width=\textwidth]{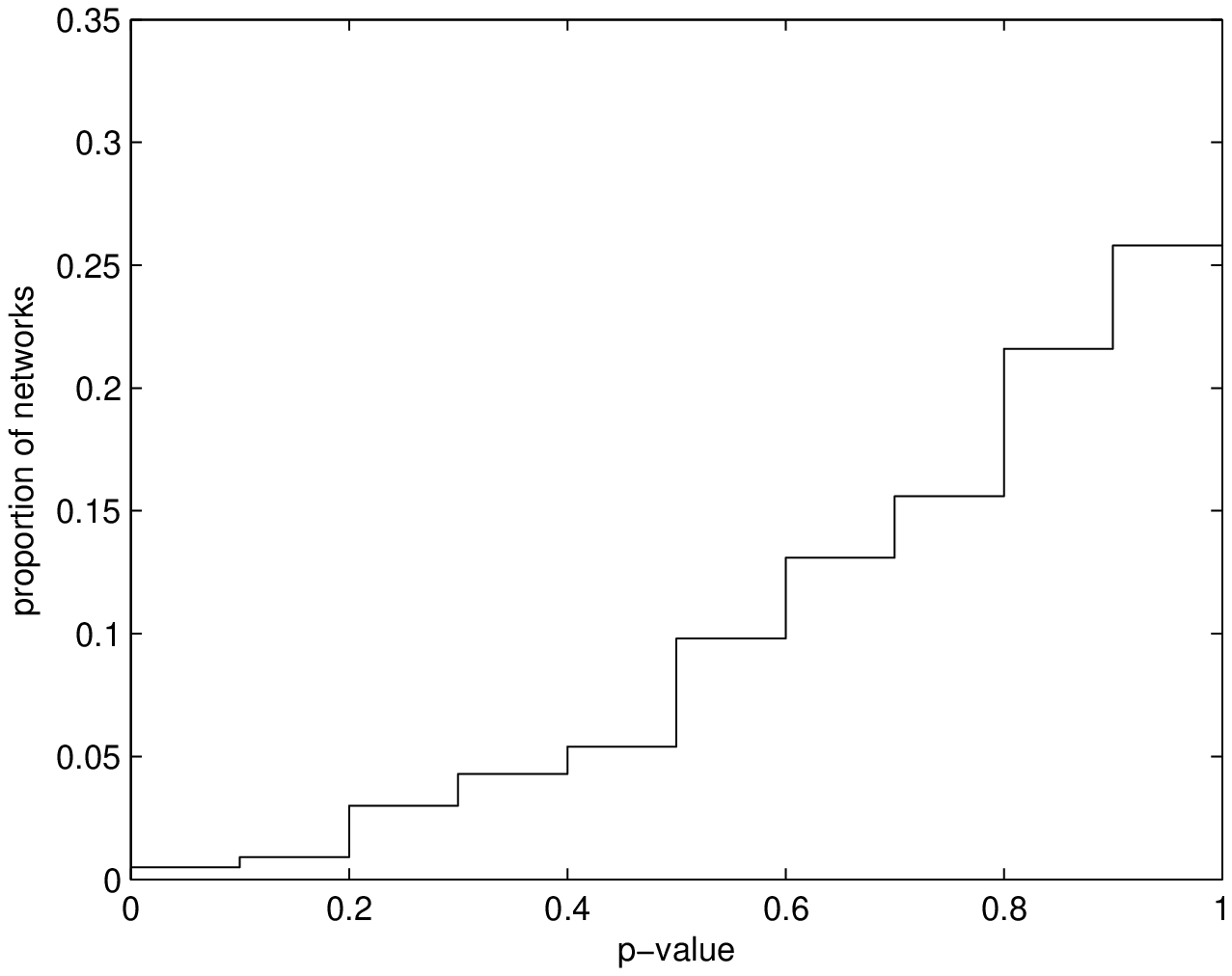}
                \label{figure8b}
        \end{subfigure}
\caption{Histograms showing the numerically derived frequencies that given p-values were achieved. The networks in the figure to the left use the parameters $p_{r}=(1/3)(\delta_{r1}+\delta_{r2}+\delta_{r3})$ and $q_{r}=(1/3)(\delta_{r2}+\delta_{r3}+\delta_{r4})$ and the networks in the figure to the right use the parameters $p_{r}=\delta_{r1}$ and $q_{r}=\delta_{r1}$, each network has $N=10^{3}$.}
\label{fig:figure9}
\end{figure}

In the numerical tests we grew a network in three phases, initially a small number of nodes with large out degree are created (the degree  must be large enough to allow $T_{l}$ to be non-zero for all $l$), then a phase of creating new nodes with a random number of out links to randomly selected nodes already in the network, finally the process described in Section \ref{description} is applied for a large number of iterations. To assess the goodness-of-fit of the results in Eqs. (\ref{gammas}) and (\ref{gammas2}) we compare the degree distribution of a simulated network of size $N$ to the distribution given by drawing $N$ values from a pseudo-random number generator adapted to output the value $k$ with probability given by Eq.(\ref{gammas2}). The degree distribution of the simulated network is then compared against the mean field prediction Eq.(\ref{gammas2}) using a suitable measure of similarity, in this case we choose the Kolmogorov-Smirnov statistic. Lastly, over a large number of trials (we chose $10^{3}$) the pseudo-random distribution is measured against the model, the p-value for this test is the proportion of trials in which the simulated data is closer to the model (i.e. a lower KS statistic) than the random data. Here we have followed the methodology of \cite{clauset}, developed for use in empirical studies where the data are not likely to be as clean as those generated in a computer simulation. The authors suggest that a p-value greater than $0.1$ is evidence enough for the model to be accepted. We ran this test for $10^{3}$ networks generated first using the pair of distributions $p_{r}=(1/3)(\delta_{r1}+\delta_{r2}+\delta_{r3})$ and $q_{r}=(1/3)(\delta_{r2}+\delta_{r3}+\delta_{r4})$ then another $10^{3}$ networks using $p_{r}=\delta_{r3}$ and $q_{r}=\delta_{r4}$. Fig.\ref{fig:figure9} shows the proportion of these trials that achieved particular p-values, while the p-value varies greatly, only a very small proportion are less than $0.1$.
\newline

\begin{figure}[h]
  \centering
 \includegraphics[width=1.1\textwidth]{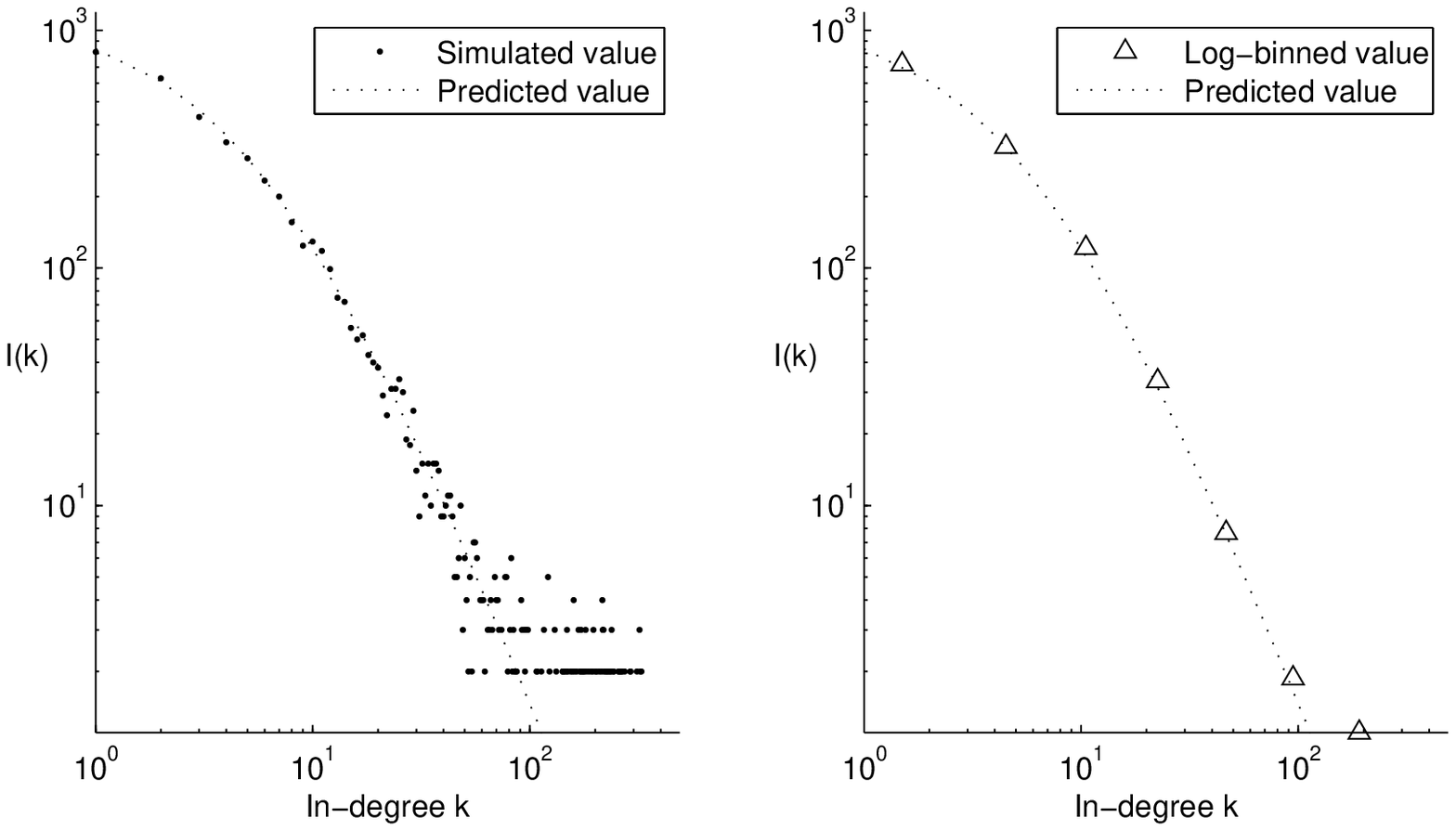}
  \caption{The number of nodes $I(k)$ of in-degree $k$ for each value of $k$, results here are taken from the simulation of the model when $p_{r}=\delta_{r3}$ and $q_{r}=\delta_{r4}$, the dotted line shows the predicted result derived from Eq.\ref{gammas}, the right hand figure shows that the log-binned values agree very well with the prediction with the exception of the largest values of $k$. This network contained $6\times 10^{3}$ nodes, the first $300$ were added randomly.}
  \label{fig:figure5}
\end{figure}

\begin{figure}[h]
  \centering
 \includegraphics[width=1.1\textwidth]{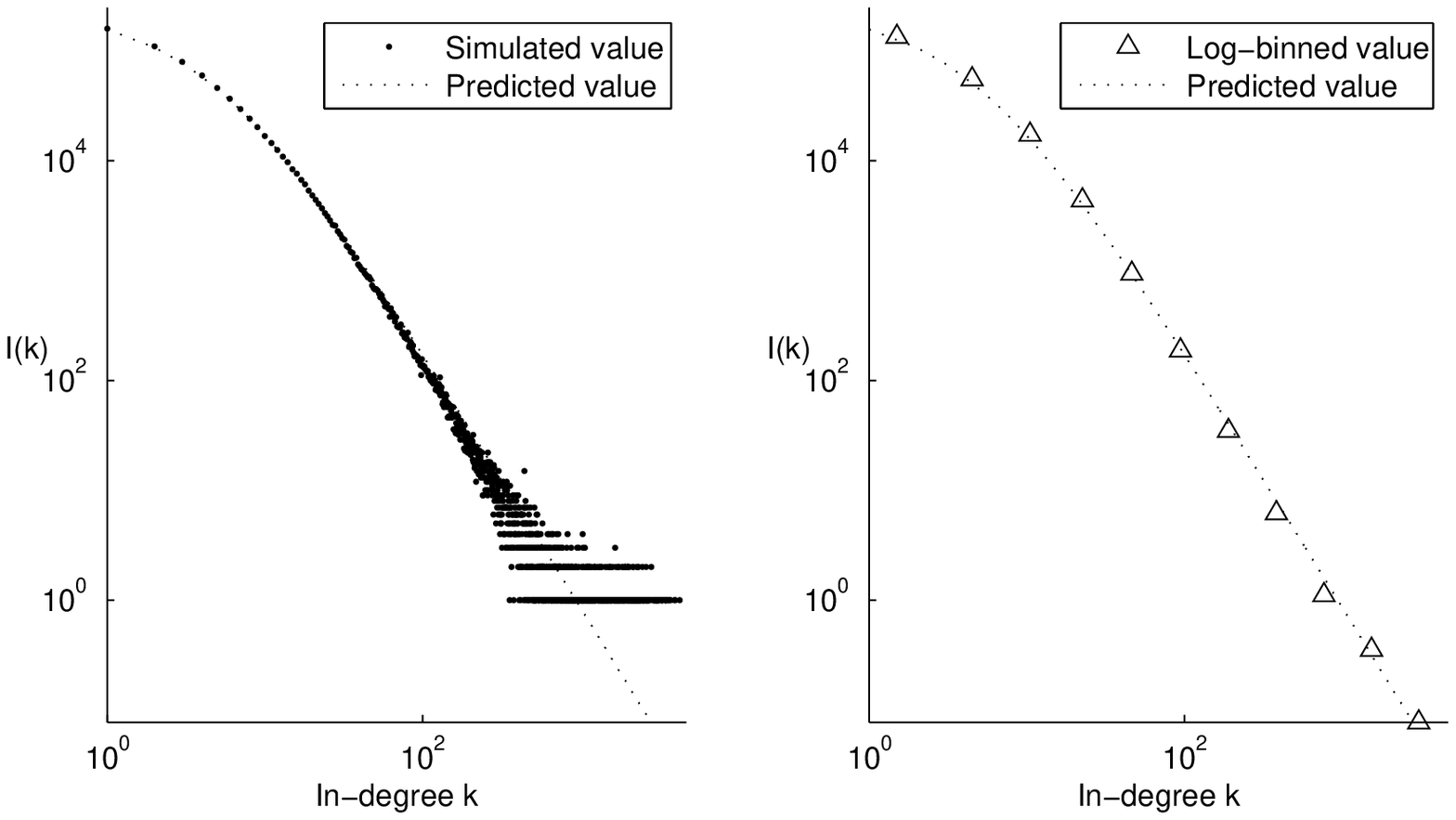}
  \caption{The in-degree from the simulation of the model when $p_{r}=(1/3)(\delta_{r1}+\delta_{r2}+\delta_{r3})$ and $q_{r}=(1/3)(\delta_{r2}+\delta_{r3}+\delta_{r4})$, the dotted line shows the predicted result given by Eq.\ref{gammas2}. This network contained $10^{6}$ nodes, the first $10^{3}$ were added randomly.}
  \label{fig:figure7}
\end{figure}

We ran the simulation for a large number of different distributions $p_{l}$ and $q_{m}$ and found that the numerical results agreed with the analytically derived formulae, Figures \ref{fig:figure5} and \ref{fig:figure7} show two typical examples. The log-binned values are the means of $I_{k}$ over a ranges of $k$ that increases logarithmically with $k$. In these examples the first bin is just the first value of $I_{1}$, the second is the mean of $I_{2}$ and $I_{3}$, the third is the next $4$ values and so on. We were able to compute the clustering coefficient only for networks no more than approximately $10^{3}$ nodes, we found that for networks where the out degree of the nodes is large the simulated result tended to be higher than the analytical result, this exposes the assumption in the analytical calculations that ambassador nodes will not be close to each other in the network. This discrepancy gets smaller as the network grows larger as one would expect.

\section{Remarks}
\label{remarks}
There are two particular strengths of this model that are worth highlighting. The first is tunability; the feature that a wide range of results for the clustering and power-law exponent can be achieved by inputting the appropriate parameter values. In the simplified model $l$ can be tuned to achieve any exponent between $2$ and $3$, by adjusting $m$ the clustering is tunable to a restricted range of values (see Fig.\ref{fig:figure6}). It is not difficult to find distributions in the full model that allow the clustering to be tuned to any value between $0$ and $1$, however as we showed in Section \ref{general} the exponent in the distribution depends on both $m$ and $l$. Tunable networks are particularly useful to study processes on networks such as epidemic spread; since the results they obtain depend largely on the topologies of the underlying networks, adjustability allows the extent of the effects of clustering and degree distribution to be analysed in greater detail \cite{epidcluster}.
The second strength of this model is its generality; the property that there are a wide range of parameter values that can be used as input to the model. As there are no restrictions on the probability distributions involved it is possible to choose those that most closely match the empirical data. A possible analysis would involve approximating the distributions $p$ and $q$ by measuring the distribution of the number of citations in the bibliography of each paper in a dataset and the distribution of citations that are also included in the bibliographies of other cited papers.

\section*{Acknowledgements}
ERC is grateful to the EPSRC for financial support.

\bibliography{bibfile2}
\bibliographystyle{model1}

\end{document}